# A Novel Watermarking Scheme for Detecting and Recovering Distortions in Database Tables

Hamed khataeimaragheh[1] and Hassan Rashidi[2]

[1]Department of Computer Engineering, Qazvin Azad University, Qazvin, Iran
`hamedkhm@yahoo.com`
[2]Department of Statistics, Mathematics and Computer Science, Allameh Tabataba'i University, Tehran, Iran
`hrashi@gmail.com;hrashi@atu.ac.ir`

## Abstract

*In this paper a novel fragile watermarking scheme is proposed to detect, localize and recover malicious modifications in relational databases. In the proposed scheme, all tuples in the database are first securely divided into groups. Then watermarks are embedded and verified group-by-group independently. By using the embedded watermark, we are able to detect and localize the modification made to the database and even we recover the true data from the database modified locations. Our experimental results show that this scheme is so qualified; i.e. distortion detection and true data recovery both are performed successfully.*

## Keywords

*Digital Watermark; Database Security; Distortion Recovery*

## 1. Introduction

With the widespread use of computers and internet, access and exchange of digital data became an extremely simple task. Since digital data can be easily duplicated and modified, there is a great deal of concern about the integrity and intellectual property protection of these data.

A new technology, known as digital watermarking, provides a promising method of protecting digital data from illicit copying and manipulation by embedding a secret code directly into the data. The embedded secret code, called watermark, can be used in various applications such as copyright protection, integrity checking, and fingerprinting.

In short, digital watermarking refers to embedding a secret imperceptible signal (watermark) in the original data. In this paper, we mainly consider digital watermarking schemes for database integrity. Generally, the digital watermarking for integrity verification is called fragile watermarking as compared to robust watermarking for copyright protection.

In a robust watermarking scheme, the embedded watermark should be robust against attacks which aim at removing the watermark or making it undetectable. While in a fragile watermarking scheme, the embedded watermark should be fragile to modifications so as to detect and localize or even recover the modifications. Most of the fragile watermarking scheme studied in the last few years, were about multimedia watermarking. Most of them focus on digital images [7, 8]; some have been extended to digital video, and audio data [4, 15].

Recently, some researchers have recognized the importance of watermarking databases and proposed some watermarking schemes to protect relational databases [2]. However, these schemes are robust schemes, which are designed for copyright protection. Though it is very important to protect the ownership of databases, sometimes, we may not care about others making copies of databases. What we care about is that the relational databases are authentic and any modifications can be detected or recovered.

This is increasingly important in many applications where relational databases are publicly available on the Internet. For example, to provide convenient access to information for users, governmental and public institutions are increasingly required to publish their data on the Internet [17]. The released data are usually in tabular form. They may be statistical data produced by Census Bureau demographic surveys and Federal agencies such as National Center for Education Statistics and Energy Information Administration; they may also be databases released by the Department of Motor Vehicles and Health Maintenance Organizations [19]. In these cases, all released data are public; what is critical for the owner of the data is to make sure that the released data are not tampered with.

We can consider another application of edge computing where databases are distributed to edge servers that perform data processing on behalf of the central server [17]. Since the edge servers may not be trusted, to ensure the relational databases are not modified by the edge servers, the central server may need to check the integrity of the relational databases regularly.

To check the integrity of relational databases, an intuitive method is to use the traditional digital signature to detect the alterations of databases [3]. A hash value is computed over a relational database and then is signed with the owner's private key. The generated signature is then appended to the database or stored separately. Though this method is very simple, there are some problems with it. First, the signature can only be used to verify whether the database has been modified or not; it cannot be used to localize and recover the modifications. Second, for a very large database, the failure of integrity verification will render the whole database useless. Finally, if there is a need to make some necessary modifications to the database, we have to compute a new signature and discard the previous one. Besides, it is computationally intensive to generate and verify the signatures. Due to these problems, it is desirable that we have a fragile watermarking scheme for relational databases so that any modifications can be detected, localized and successfully recovered. In this way, even if some part of a relational database has been altered, after localizing these modifications, and then we only need to repair the modified data by recovering the true data from modified locations and in such way we ensure that the whole of the data set is still authentic and reliable. In addition, because the fragile watermarking scheme is private key based, its computational cost is obviously less than that of digital signature schemes.

In this paper, we proposed a fragile watermarking scheme to verify the integrity of a relational database and recover the true database. In the proposed scheme, all tuples in a relational database are first divided into groups by using a secret key. After that Watermarks are embedded in all groups and then watermark verification done independently in each group. The embedded watermarks cannot only detect but also localize the modifications made to the database, and even recover the true data from modified cells in database tables. Since the watermarking process will inevitably introduce small distortions to a relational database, we assume that the relational database to be watermarked has numerical attributes which tolerate small changes.

The rest of this paper is organized as follows. Section 2 gives an overview of related work. Section 3 explains in detail our proposed fragile watermarking scheme, including watermark embedding, watermark detection and true data recovery. Security analysis of the scheme is provided in Section 4. Section 5 concludes this paper with summaries and suggestions for future work.

## 2. RELATED WORK

In recent years, there has been relatively few works on watermarking relational databases. Agrawal and Kiernan [2] present a robust watermarking scheme for databases. According to an embedding key, some bits of some attributes of some tuples are modified to embed watermark bits. Li et al. [11, 12] further extend this scheme to embed multiple bits information instead of one bit information as in Agrawal and Kiernan's scheme into a relational database so that potential illegal distributors can be tracked. Also, this scheme is claimed to be more robust since false negative and false positive detection rates are bounded.

Sion et al. [18] present a different approach to robust watermarking scheme for databases. In their scheme, all tuples are securely sorted and divided into non-intersecting subsets. A single watermark bit is embedded into some tuples of a subset by modifying the distribution of tuple values. Watermark bits are embedded repeatedly and an error control coding scheme is employed to recover the embedded bits. This scheme is claimed to be robust against attacks such as data resorting and data transformations.

Another scheme is proposed by Gross-Amblard [9], where database instances with bounded degree Gaifman graph are watermarked while local queries are preserved. They also show that the difficulty of query-preserving watermarking is linked to the informational complexity of sets defined by queries. However, they do not analyze the robustness of their scheme, which is a very important property of watermarking schemes for copyright protection.

Devanbu et al. [6] and Pang and Tan [17] present schemes that are based on the Merkle Hash Tree, where each tuple is treated as a leaf node and a verifiable B+ tree is constructed by adding signed digest for very attribute and for each leaf node recursively until the root node is reached. Verification objects are created for query operations to authenticate query results. Though these schemes can detect any modifications, they cannot localize the modifications. In addition, they have extra overhead for storing and maintaining the tree.

All current robust watermarking schemes for relational databases are designed for copyright protection. Guo et al. [1] present a fragile watermarking scheme that cannot only detect but also localize the modifications made to the database, it was the most complete scheme which has been introduced yet. In this scheme all tuples in a relational database are first divided into groups then by using a hash function Watermarks are embedded and verified group by group independently according to some security parameters. The embedded watermarks can detect and localize the modifications made to the database. In the worst case, the modifications can be narrowed down to tuples in a group.

All current watermarking schemes for relational databases are designed for copyright protection or localizing the distortions made to the database tables. But none of them can recover the true data from the modified cells in database tables.

In contrast, our scheme is a novel fragile watermarking scheme that cannot only detect and localize the modifications made to the database tables, but also recover the true data from modified cells of tables.

## 3. THE PROPOSED FRAGILE WATERMARKING ALGORITHM

This section introduces our proposed fragile watermarking scheme which is able to detect and localize any malicious modification made to a relational database and also it can recover the true data from modified cells. We assume that all attributes of the relational database are numeric and the database can tolerate minor distortions introduced by watermark embedding. Though the order of tuples can be changed arbitrarily, we assume without loss of generality that the order of attributes is not changed.

## 3.1. Watermark Embedding

Suppose there is a relational database which has a primary key P and y attributes, denoted by T(P,A1,A2, . . . ,Ay). All attributes are numeric. Since watermarking will inevitably introduce small distortions to attribute values, it is assumed that each attribute value can tolerate modifications of at least two least significant bits. Table 1 shows the parameters along with their description that are used in this paper.

Table 1: Parameters and their description

| Parameter | Description |
| --- | --- |
| **y** | Number of attributes in the relation |
| **w** | Number of tuples in the relation |
| **g** | Number of groups in the table |
| **v** | Average number of tuples in a group |
| **$G_k$** | The k-th group |
| **$r_i.A_j$** | The j-th attribute of the i-th tuple in a group |
| **K** | Watermark embedding key |
| **$W^j_1$** | Watermark embedded in the j-th attribute of all tuples in a group |
| **$W^i_2$** | Watermark embedded in all attributes of the i-th tuple in a group |
| **$W^{j*}_1$** | Original watermark extracted from p(j)-th attribute of all tuples in a group |
| **$W^{i*}_2$** | Original watermark extracted from p(j)-th attribute of all tuples in a group |
| **$V^j_1$** | Watermark verification result for $W^j_1$ |
| **$V^i_2$** | Watermark verification result for $W^i_2$ |

The watermark embedding algorithm is shown in Algorithms 1 and 2. All tuples are first divided into groups according to the number of groups g, the hash value of a embedding key K, and their primary key. For the security reason, the embedding key should be selected from a large enough key space so that it is computationally infeasible for an attacker to guess the key. Only the one who has the knowledge of K and g can determine the group that the tuples belong to. The tuples in each group are then sorted based on their primary key. The grouping and sorting operations are very important in our algorithm.

Since tuples in a relational database are independent, it is very important to enforce some relationship between them so that the embedded watermarks and the extracted watermarks can be synchronized. Note that grouping and sorting do not change tuples' physical positions in the table.

In watermark embedding, watermark detection and true data recovering, we ignore the least two significant bits of all attributes except the primary key when computing hash values.

Algorithm 1: The pseudo code for Watermark embedding

```
1: For  i =1 to  w  do
2:   hk_i =hash(k,r_i.p)
3:   k =hk_i  mod  g
4:   r_i  →  G_k
5: End for
6: For  k=1 to g  do
7:   sort tuples  in  G_k  according  to
     Their primary   key
8:   embed watermarks to all tuples in G_k
     //see the algorithm 2
9: End for
```

Next, watermarks are embedded into each group independently. In each group, there are two kinds of watermarks:

- Attribute watermark $W_1$ which consists of y watermarks of length v.
- Tuple watermark $W_2$ which consists of v watermarks of length y.

Accordingly, the embedding subroutine mainly consists of two parts: attribute watermark embedding and tuple watermark embedding. In each part, the watermarks to embed are first extracted from a one-way hash value.

For attribute watermarks, the hash value is generated according to a message authentication code and the same attribute of all tuples in the group, while for tuple watermarks, it is formed according to the same message authentication code and all attribute values of the same tuple. For generating the hash value, we use XOR operator as a hash function instead of other miscellaneous hash function. This operator has heredity nature and therefore our scheme is able to recover the true data from the modified cells of database table.

The watermark embedding is very simple. For any value $r_i.A_j$, in attribute watermark embedding, the least significant bit of $r_i.A_{p(j)}$ is set to i-th bit of $W^j_1$. By using this function our scheme is able to embed the extracted attribute watermark from j-th column of group ($W^j_1$ watermark), in another column of group (p(j)-th column of group). This solution is required for true data recovery process. The p(j) is shown in the algorithm 3. In the scheme [1], there is not such a function. In that scheme generated attribute watermarks from j-th attribute values of groups (j-th column values of group) is embedded in the same attribute values (j-th column values of table) of groups.

In tuple watermark embedding, the next least significant bit of $r_i.A_j$ is set to the j-th bit of $W^i_2$. In this way, the embedded watermarks actually form a watermark grid, which helps to detect, localize and recover modifications.

### 3.2. Watermark Detection

The watermark detection algorithm is shown in Algorithms 4 and 5. To verify the integrity of a database relation, we need to know K and g. As in watermark embedding, all tuples are divided into groups and each group is verified independently. For each group, we construct two verification vectors $V_1 = (V^1_1, V2_1... V^y_1)$ and $V_2 = (V^1_2, V^3_2... V^v_2)$ where $V^j_1$ denotes the verification result for attribute watermark $W^j_1$, and $V^i_2$ denotes the verification result for tuple watermark $W^i_2$.

Each element of the vectors is either true or false depending on whether the embedded watermark matches the related extracted watermark. Since $V_1$ is for $W_1$ and $V_2$ is for $W_2$, to get $V^j_1$ (j ∈ [1, y]), the watermark extracted from the least significant bits of all tuples' j-th attribute and the watermark constructed from the hash value of the message authentication code and the j-th attribute of all tuples in a same group are compared. If the two matches, $V^j_1$ is true, otherwise, it is false. Likewise, to get $V^i_2$ (i ∈ [1, v]), the watermark extracted from the next least significant bits of the i-th tuple's all attributes and the watermark constructed from the hash value of the message authentication code and all attributes of the i-th tuple are compared. If the two matches, $V^i_2$ is true, otherwise, it is false. From the two vectors $V_1$ and $V_2$, we can detect, localize and characterize any modifications made to tuples in a group.

From the watermark embedding and detection algorithms, we can easily see that the computational complexity of both algorithms is in the order of the number of tuples in the relational database; that is, the complexity is O(w).

Algorithm 2: the pseudo code for Watermark embedding in Gk

```
1:  For j=1 to y  do
2:    H^j_1 = XOR(k , r_1.A_j , r_2.A_j ,…, r_v.A_j)
      //exclude the least two significant bits
      // of all values
3:    W^j_1=extractbits(h^j_1,v) //see algorithm 5
4:    For i=1 to v do
5:      W^j_1(i) ⟶ least significant bit of r_i.A_{p(j)}
          // p(j) is shown in algorithm 3
6:    End for
7:  End for
8:  For i = 1 to v do
9:    h^i_2 =hash (k, r_i.A_1, r_i.A_2 ,…, r_i.A_y)
      //exclude the least two significant bits
      //of all values
10:   W^i_2= extractbits(h^i_2,y)    //see algorithm 6
11:   For j=1 to y do
12:     W^i_2(j) ⟶ next least significant bit of  r_i.A_j
13:   End for
14: End for
```

Algorithm 3: The pseudo code of p (j)

```
1: p(j)= ( (k+j) mod (y-1) ) +1
   // return a number between [1..y]
```

Algorithm 4: The pseudo code for Watermark verification

```
1: for  i =1 to  w  do
2:    hk_i = hash (k,r_i.p)
3:    k = hk_i  mod  g
4:    r_i ⟶ G_k
5: end   for
6: for  k =1 to g  do
7:   Sort tuples in G_k according to their
     primary key  value
8:   Verify the integrity of G_k
     //see algorithm 5
9: End for
```

Algorithm 5: the pseudo code for Verification the authenticity of Gk

```
1:  For j=1 to y do
2:    h^j_1 = XOR(k , r_1.A_j , r_2.A_j , … , r_v.A_j)
      //exclude the least two significant bits
      //of all values
3:    W^j_1 = extractbits (h^j_1, v)
      //see algorithm 6
4:    For i=1 to v do
5:      W^{j*}_1(i) ← least significant bit of r_i.A_{p(j)}
6:    End for
7:    If W^{j*}_1 ≠ W^j_1 then V^j_1 = false
8:    Else V^j_1 = true
9:    End if
10: End for
11: For j = 1 to y do
12:   If (V^j_1 = false) and (V^{p(j)}_1 = false)
          then V^j_1 = true
13:   End if
14: End for
15: For i=1 to v do
16:   h^i_2 = hash (k, r_i.A_1, r_i.A_2 ,…, r_i.A_y)
      //exclude the least two significant
      // bits of all values
17:   W^i_2 = extractbits(h^i_2, y)
      //see algorithm 6
18:   For j=1 to y do
19:     W^{i*}_2(j) ← next least significant bit of r_i.A_j
20:   End for
21:   If w^{i*}_2 ≠ w^i_2 then V^i_2 = false  else V^i_2 = true
22:   End if
23: End for
```

Algorithm 6: The pseudo code for ExtractBits(H,L)

```
1: If length (H) ≥ L then
2:   W = concatenation of L most
       significant bits from H
3: Else m = L – length(H)
4:   W = concatenation of H and extractbits (H,M)
5: End if
6: Return W
```

### 3.3. Localization of the Modifications

The ability to localize the modifications is one of our scheme advantages. Since the tuples are grouped based on the number of groups g and the hash value of the embedding key and their primary key, changing the order of tuples does not affect the embedded watermarks.

It is reasonable that the proposed scheme accept such modifications. In our scheme, a single value modification can be detected, localized and recovered easily. In the following, we will illustrate how to detect a single modification.

Assume the tuples of a relational database are already divided into groups and the tuples in each group are sorted. Figure 1 shows a group which contains 4 tuples and 4 attributes. Suppose $r_2.A_3$ is altered, all relevant watermarks $W^3_1$ and $W^2_2$, which are hash functions of $r_2.A_3$, also change.

Then for all k ($1 \leq k \leq 4$), $V^k_1$ is true except $V^3_1$. For all q ($1 \leq q \leq 4$), $V^q_2$ is true except $V^2_2$. From this result, we can easily see that $r_2.A_3$ has been altered. Figure 1 shows the related verification result. The blue line denotes successful verification while the red line denotes failed verification. This example shows the detection and localization of a non-primary key modification. If the modified value is a primary key, the modification can still be located, but it can only be located to one or two groups.

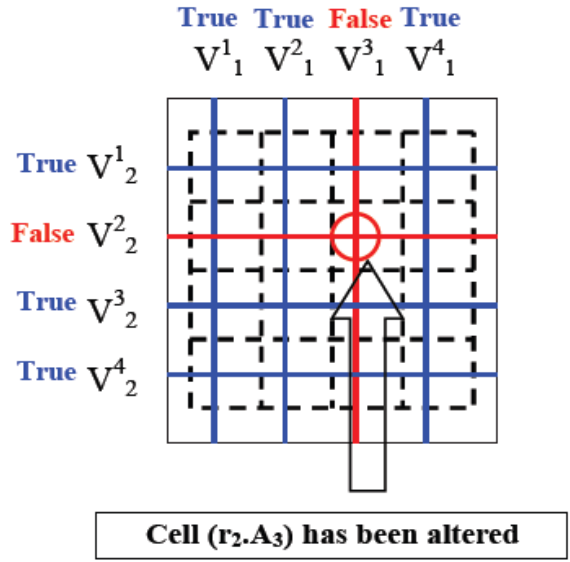

Figure 1: An illustration of verification results in a cell

### 3.4. True Data Recovery

Other advantage of our proposed scheme is the ability to recover the true data from modified cells. After detection and localization the modified cells, our scheme can recover the true data. As mentioned, in attribute watermark embedding we use XOR operator as a hash function to generate attribute watermark code. The input values of this operator to generate j-th attribute watermark ($w^j_1$), include the j-th attribute values of group and output value of this operator was embedded in p(j)th attribute values of group therefore by altering one cell value in j-th attribute value of a tuple in group, our scheme can extract j-th original attribute watermark of group($w^{j*}_1$) and by using  heredity nature of XOR operator it can recover the original single value of modified cell which is participates in original attribute watermark  generation. The true data recovery process is shown in algorithm 7.

## 4. SECURITY ANALYSIS

In this section we evaluate the security of proposed scheme to recover the true data from single modified cell in a database tables. In this evaluation we apply the proposed scheme as an application to a database table which has been distorted. We run our scheme on the modified table in order to recover the true data from single modified cell. This work is done on several tables with the parameter (v=10, 30, 50), (y=10, 20, 30, 40, 50).

Algorithm 7: The pseudo code for recovering true data from modified cell

```
1:  For j= 1 to y do
2:    If V^j_1 = false then modifiedcolumn = j
3:  End for
4:  For i=1 to v do
5:    If V^i_2 = false then modifiedrow = i
6:  End for
7:  S = k     // k is the secret key
8:  For i= 1 to v do
9:    If i ≠ modifiedrow then
10:     S = S xor r_i.A_modifiedcolumn
        //exclude the least two significant
        // bits of all values
11:   end if
12: end for
13: For i=1 to v do
14:   W^{modifiedcolumn*}_1(i) ← least significant bit
        of r_i.A_{p(modifiedcolumn)}
15: End for
16: r_modifiedrow.A_modifiedcolumn = (w^{modifiedcolumn*}_1) xor (S)
```

The proposed scheme was done 10,000 times for each table, and we calculated the recovery failure probability for each table by dividing the number of failure to 10,000. The result is shown in Figure 2. According to that, we find out two main conclusions:

- Recovery Failure probability in all modified table is very teeny as in worst case the recovery failure probability for a table (v=10 and y=10) is about 0.001.

- With the growth of table tuples (v) and tuple attributes (y) the recovery failure probability decreases.

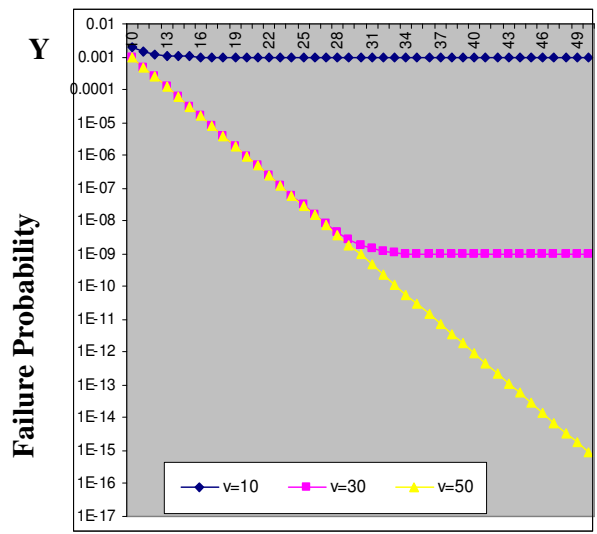

Figure 2: Failure probabilities for a single value modification.

10International Journal of Database Management Systems ( IJDMS ) Vol.2, No.3, August 2010## 5. CONCLUSIONS

In this paper, we proposed a fragile watermarking scheme for relational databases. The watermarks are embedded into a relational database on the group basis under the control of a secure embedding key. The embedded watermarks form a watermark grid which can detect and localize any modifications made to the database and also be able to recover true data from modified cells. Experimental results showed that proposed scheme is secure and true data recovery failure probability is very teeny.

Security analysis showed that it is very difficult for an attacker to modify the database without affecting the embedded watermarks, and the security upper bound was given. Future work will focus on designing a watermarking scheme that can embed watermarks to non-numeric attributes. For this purpose we can choose two solutions. The first solution is to reform the structure of hash function so it can accept non numeric inputs. The second solution would be another mechanism instead of using a hash function.